\documentstyle[medium98,12pt,epsbox]{article}

\newcommand{\vct}[1]{\mbox{\boldmath{$#1$}}}
\newcommand{\brkt}[3]{\langle#1|#2|#3\rangle}
\newcommand\beq{\begin{equation}}
\newcommand\eeq{\end{equation}}
\newcommand\beqa{\begin{eqnarray}}
\newcommand\eeqa{\end{eqnarray}}

\begin{document}                      
\title{
The Landau-Migdal Parameters, $g'_{{\rm NN}}$ and \ $g'_{{\rm N}\Delta}$\\
and Pion Condensations
}
\author{
T.~Suzuki$^{1,3}$, 
H.~Sakai$^{2,3}$, 
and 
T.~Tatsumi$^4$}
% Replace the following addresses to your address(es)
\address{$^1$Department of Applied Physics, Fukui University, Fukui
910-8507, Japan}
\address{$^2$Department of Physics, University of Tokyo, 
   Tokyo 113-0033, Japan}
\address{$^3$RIKEN, 2-1 Hirosawa, Wako-shi, Saitama 351-0198, Japan}
\address{$^4$Department of Physics, Kyoto University, 
   Kyoto 606-8502, Japan}

\maketitle                             
\abstracts{
The Landau-Migdal parameters for nucleon-nucleon($g'_{{\rm NN}}$) and nucleon-$\Delta(g'_{{\rm N}\Delta})$ couplings are estimated from recent experimental data on the giant Gamow-Teller state of $^{90}$Nb.
The observed quenching of the GT strength by 10\% provides $0.16 < g'_{{\rm N}\Delta} <0.23$ for $g'_{\Delta\Delta}<1.0$, while the excitation energy $g'_{{\rm NN}} \sim $ 0.59.  
The critical density of pion condensations is much lower than those previously discussed with the universality ansatz.
}

\section{Introduction}
The Landau-Migdal parameters for nucleon-nucleon($g'_{{\rm NN}}$), nucleon-$\Delta(g'_{{\rm N}\Delta})$ and $\Delta-\Delta(g'_{\Delta\Delta}$) couplings play a crucial role in the study of spin-dependent structure of nuclei. In particular, they dominate pion condensations in high density nuclear matter\cite{t}. 
Experimentally, however, their values are not known well yet.
Since experimental information was very limited, they were frequently estimated by assuming the universality, $g'_{{\rm NN}}=g'_{{\rm N}\Delta}=g'_{\Delta\Delta}$\cite{t,r,alb,bm}. For example, excitation energies of spin-isospin dependent states  provided the  parameters to be $0.6\sim 0.8$\cite{t}.
It is shown that for these values pion condensations are expected very hardly even in high density nuclear matter\cite{t}. 

In 80's, the giant Gamow-Teller(GT) states were observed\cite{o}.
Their excitation energy and strength provide two independent experimental values for three Landau-Migdal parameters. 
The strength, however, was not fully determined. At least about 50\% of the GT sum rule value was observed, but it was not clear where the rest of the strength exists. Nevertheless, if one assumes that about 50\% of the strength is quenched owing to the coupling of the particle-hole states with the $\Delta$-hole states, the universality ansatz with $0.6\sim0.9$\cite{s1,s2} explains well both excitation energy and the strength. As a result, the universality ansatz was widely used for a long time.

Recently, Wakasa et al.\cite{w} have determined the GT strength of $^{90}$Zr experimentally.
They have observed 0.90 of the GT sum rule value with the statistical error, $\pm$0.05. This fact implies that the universality ansatz does not hold.
The purpose of the present paper is to estimate the values of $g'_{{\rm NN}}$ and $g'_{{\rm N}\Delta}$ from the recent experimental data, and to discuss pion condensations. 
We will show that the quenching of the strength strongly depends on the value of $g'_{{\rm N}\Delta}$ and weakly on that of $g'_{\Delta\Delta}$. Consequently, we can determine the value of $g'_{{\rm N}\Delta}$ from the magnitude of quenching to be about 0.2. Moreover, because of this small value, the excitation energy depends almost on the value of $g'_{{\rm NN}}$ only. This fact yields $g'_{{\rm NN}}$ to be about 0.6, indicating that the universality ansatz\cite{r,bm} does not hold. The small $g'_{{\rm N}\Delta}$ makes the critical density of pion condensations very low, compared with that discussed with the universality ansatz.

\section{Model}

\subsection{The GT states in the nucleon space}

First we discuss the GT states within the nucleon space\cite{s3}. The hamiltonian is assumed to be
\begin{equation}
H = H_0 + H_1 + V_{{\rm c}},
\end{equation}
where $V_c$ and $H_0$ denotes the Coulomb part and the spin and isospin-independent part of the hamiltonin, respectively. The spin and isospin dependent part is given by\cite{bm,s3}
\begin{equation}
H_1 = -\sum_{i=1}^A \xi_i\vct{l}_i\cdot \vct{\sigma}_i + \frac{1}{2}\kappa_\tau\sum\vct{\tau}_i\cdot \vct{\tau}_i + \frac{1}{2}\kappa_{\tau\sigma}\sum(\vct{\tau}_i\cdot\vct{\tau}_j)(\vct{\sigma}_i\cdot\vct{\sigma}_j).
\end{equation}
For this hamiltonian, the excitation energy of the isobaric analogue state(IAS) is expressed by
\begin{equation}
E_{{\rm IAS}} = E_\pi + \epsilon_\pi + 4\kappa_\tau T_0, \label{ias}
\end{equation}
where $E_\pi$ stands for the energy of the parent nucleus, while $\epsilon_\pi$ and $T_0$ the unperturbed energy and the isospin of IAS, respectively. 
By taking experimental values of $E_{{\rm IAS}}, E_\pi$ and $\epsilon_\pi$ of Eq.(\ref{ias}), we can estimate the strength of the spin-isospin interaction as,
\begin{equation}
A\kappa_\tau = 23.7 {\rm MeV}\label{iso}
\end{equation}
for $^{90}$Zr\cite{s3}. The excitation energy of the GT state is given by
\begin{equation}
E_{{\rm GTS}} = E_{{\rm IAS}} + \frac{2}{3T_0}\brkt{\pi}{\sum\xi_i\vct{l}_i\cdot\vct{\sigma}_i}{\pi} - 4( \kappa_\tau - \kappa_{\tau\sigma})T_0,\label{gts}
\end{equation}
where the second term denotes the expectation value of the spin-orbit force as to the ground state of the parent nucleus. Since we know the values of the first three quantities experimentally and have determined that of $\kappa_\tau$ in Eq.(\ref{iso}), we obtain the strength of the spin-isospin interaction as,
\begin{equation}
A\kappa_{\tau\sigma} = 19.8 {\rm MeV},\label{spin}
\end{equation}
for $^{90}$Zr\cite{s3}.
In other words, the excitation energy of the GT state in $^{90}$Nb is reproduced by the strength of Eq.(\ref{spin}).

\subsection{The GT states in the nucleon and $\Delta$ space}

Next we discuss the GT states using the Landau-Migdal parameters. We assume the spin-isospin interaction in the quark model\cite{s1,s2},
\begin{equation}
V = (V_{{\rm NN}} + V_{{\rm N}\Delta} + V_{\Delta\Delta})V_q,
\end{equation}
where we have defined
\begin{equation}
V_q = \frac{1}{2}\left(\frac{f_\pi}{m_\pi}\right)^2\sum_{i,j}^A\delta(\vct{r}_i - \vct{r}_j)\{(\vct{\tau}_i\cdot\vct{\tau}_j)(\vct{\sigma}_i\cdot\vct{\sigma}_j)\}_q
\end{equation}
with
\begin{eqnarray}
(\sigma\tau)_q = \sum_{i=1}^3 \sigma^{(i)}\tau^{(i)}, \ \ \ \left(\frac{f_\pi}{m_\pi}\right)^2 = 392 \ {\rm MeV\cdot fm}^3.\nonumber
\end{eqnarray}
The relationship between the above and Landau-Migdal parameters is given by
\begin{eqnarray}
g'_{{\rm NN}} = \left(\frac{5}{3}\right)^2V_{{\rm NN}}, \ \ \ g'_{{\rm N}\Delta} = \frac{10\sqrt{2}}{3}\frac{f_\pi}{f_\Delta}V_{{\rm N}\Delta}, \ \ \ g'_{\Delta\Delta} = 8\left(\frac{f_\pi}{f_\Delta}\right)^2V_{\Delta\Delta},
\end{eqnarray}
where $f_\pi/f_\Delta$ is $\sqrt{25/72}$ in the quark model, while 1/2 in the Chew-Low model. We solve the RPA equation with the above force in the nucleon and $\Delta$ space. In order to reproduce the excitation energy of the GT state, finally we obtain the relationship between the Landau-Migdal parameters and the strength of the spin-isospin interaction of the previous subsection\cite{s4},
\begin{equation}
A\kappa_{\tau\sigma} = g'_{{\rm NN}}\left(\frac{f_\pi}{m_\pi}\right)^2\rho_0\gamma \ \frac{1 + \beta g'_{\Delta\Delta}\left\{1 - (g'_{{\rm N}\Delta})^2/(g'_{\Delta\Delta}g'_{{\rm NN}})\right\}}{1 + \beta g'_{\Delta\Delta}},\label{kg}
\end{equation}
where in the quark model $\beta$ is given by
\begin{equation}
\beta = \frac{64}{25\epsilon_\Delta}\left(\frac{f_\pi}{m_\pi}\right)^2\rho_0\gamma\left(1 + \frac{Z - N}{2A}\right).
\end{equation}
In the above equation, $\rho_0$ denotes the nuclear matter density, 
and $\epsilon_\Delta$ the unperturbed energy of the $\Delta$-hole states,
\begin{equation}
\rho_0 = 0.17 {\rm fm}^{-3}, \ \ \epsilon_\Delta = 294 {\rm MeV},
\end{equation}
and $\gamma$ stands for the attenuation factor for nuclear surface effects.
When we employ the Skyrme III force for $^{90}$Zr, it is given by $\gamma = 0.511$ for the $g$ orbit\cite{s1}. In the Chew-Low model we obtain the same equation as Eq.(\ref{kg}), but replacing $\beta$ by $\beta_1$,
\begin{equation}
\beta_1 = \frac{25}{18}\beta.
\end{equation}

The GT strength in the present model is given by
\begin{eqnarray}
|\brkt{{\rm GT}}{\sum\tau_-\sigma_-}{0}|^2 = Q\sum_{ph}|\brkt{p}{\tau_-\sigma_-}{h}|^2,
\end{eqnarray}
where $Q$ represents the quenching factor of the strength. In the quark model it is written as\cite{s4}
\begin{eqnarray}
Q = \left\{\frac{1 + \beta(g'_{\Delta\Delta} - g'_{{\rm N}\Delta})}{1 + \beta g'_{\Delta\Delta}}\right\}^2.\label{qq}
\end{eqnarray}
In the Chew-Low model, we have\cite{s4}
\begin{equation}
Q = \left\{\frac{1 + (\beta_1g'_{\Delta\Delta} - \beta_2g'_{{\rm N}\Delta})}{1 + \beta_1g'_{\Delta\Delta}}\right\}^2, \ \ \ \beta_2 = \frac{5}{3\sqrt{2}}\beta.\label{clq}
\end{equation}

\section{Estimation of $g'_{{\rm NN}}$ and $g'_{{\rm N}\Delta}$}

As mentioned before, the quenching factor of the GT strength has been observed to be 0.90 in $^{90}$Zr\cite{w}. Inserting this value into Eqs.(\ref{qq}) and (\ref{clq}), $g'_{{\rm N}\Delta}$ is expressed as a function of $g'_{\Delta\Delta}$. We obtain in the  quark model
\begin{eqnarray}
g'_{{\rm N}\Delta} = 0.18 + 0.05\ g'_{\Delta\Delta},\label{gq}
\end{eqnarray}
while in the Chew-Low model
\begin{equation}
g'_{{\rm N}\Delta} = 0.16 + 0.06\ g'_{\Delta\Delta}.\label{gcl}
\end{equation}
The value of $g'_{\Delta\Delta}$ is not known, but it may be reasonable to assume $g'_{\Delta\Delta}$ $< 1.0$\cite{r,o}. Then we have
\begin{eqnarray}
0.18 < g'_{{\rm N}\Delta} &<& 0.23 \ \ ({\rm quark\ model}),\label{ndq}\\
0.16 < g'_{{\rm N}\Delta} &<& 0.22 \ \ ({\rm Chew-Low\ model})\label{ndcl}.
\end{eqnarray}
Both models give $g'_{{\rm N}\Delta}$ to be about 0.2, which is small, compared with the one obtained in the universality ansatz.

The value of $g'_{{\rm NN}}$ is estimated by using Eqs.(\ref{spin}) and (\ref{kg}). For $g'_{\Delta\Delta}<1.0$ and Eqs.(\ref{ndq}) and (\ref{ndcl}), we obtain
\begin{eqnarray}
0.591 < g'_{{\rm NN}} < 0.594.\label{nn}
\end{eqnarray}
Thus the value of $g'_{{\rm NN}}$ is well fixed to be about 0.59. This is due to the fact that because $g'_{{\rm N}\Delta}$ is small, the excitation energy of the GT state is dominated by $g'_{{\rm NN}}$.
Eqs.(\ref{ndq}) and (\ref{ndcl}), and (\ref{nn})show that the universality ansatz does not hold. We note that if we assume the universality, the excitation energy of GT state in $^{90}$Nb provides the Landau-Migdal parameter, 0.75, yielding the quenching factor, 0.60\cite{s1,s2}.

\section{Pion condensation in nuclear matter}

The interaction among particle- and $\Delta$- hole states in the
spin-isospin channel should be dominated by the $p$-wave coupling with
pions at higher momentum, which, in turn, may give rise to the
softening of the spin-isospin sound mode, pion condensation. This possibility
has been extensively studied\cite{t}, but it is still a long-standing problem
in nuclear physics since the critical density is {\it sensitive} to the 
values of the Landau-Migdal parameters. We reexamine the critical
density by the use of
the results given in  Eqs.~(18) and (21) without the universality ansatz 
usually adopted in many papers\cite{t}. In the following we use
the Chew-Low value for the 
$\pi N\Delta$-coupling constant, $f_{\Delta}\sim 2f_\pi$.

Consider the pion propagator with $({\bf k}, \omega)$ in matter,
\beq
D^{-1}_{\pi}({\bf k},\omega)=\omega^2-m_\pi^2-k^2-\Pi({\bf k},\omega;\rho)
\eeq
with the self-energy term $\Pi$, which represents the interactions of
pions 
with nucleon and $\Delta$ -hole states, 
$
\Pi=\Pi_N+\Pi_\Delta.
$
Then the threshold conditions for pion condensations are given by 
\beq
D^{-1}_\pi(k_c, \omega=0; \rho_c)=0, \quad 
\partial D^{-1}_\pi/\partial k|_{k=k_c}=0
\eeq
with the critical density ($\rho_c$) and momentum ($k_c$) 
for neutral-pion condensation, and
\beq
D^{-1}_\pi(k_c, \omega=\mu_\pi^c; \rho_c)=0, \quad 
\partial D^{-1}_\pi/\partial k|_{k=k_c}=0, \quad
\partial D^{-1}_\pi/\partial \omega|_{\omega=\mu_\pi^c}=0
\eeq
with the critical chemical potential ($\mu_\pi^c$) for charged-pion 
condensation.

\subsection{Neutral pion $(\pi^0)$ condensation}

First, we consider the neutral-pion condensation. In this case the 
$s$-wave pion-baryon interaction should be small, so that only
the $p$-wave pion-baryon interaction is relevant for the self-energy, 
$\Pi=\Pi_N^P+\Pi_\Delta^P$; 
\beqa
\Pi_N^P&=&-k^2U^{(0)}_N\left[1+(g'_{\Delta\Delta}-g'_{N\Delta})U^{(0)}_\Delta\right]/D\nonumber\\
\Pi_\Delta^P&=&-k^2U^{(0)}_\Delta\left[1+(g'_{NN}-g'_{N\Delta})U^{(0)}_N\right]/D
\eeqa
with
\beq
D=1+g'_{NN}U^{(0)}_N+g'_{\Delta\Delta}U^{(0)}_\Delta
+(g'_{NN}g'_{\Delta\Delta}-g^{'2}_{N\Delta})U^{(0)}_NU^{(0)}_\Delta,
\eeq
where $U^{(0)}_N$ and $U^{(0)}_\Delta$ are the polarization functions, 
represented in terms of the standard Lindhard functions for
nucleons $(L_N)$ and $\Delta's$ $(L_\Delta)$\cite{mig}, respectively,
\beq
U^{(0)}_N\equiv\left(\frac{f_\pi
\Gamma_\Lambda}{m_\pi}\right)^2L_N, \quad 
U^{(0)}_\Delta\equiv 
\left(\frac{f_\Delta\Gamma_\Lambda}{m_\pi}\right)^2L_\Delta.
\eeq
Here we have introduced the form-factor for the $p$-wave coupling vertex,
\beq
\Gamma_\Lambda=\frac{\Lambda^2-m_\pi^2}{\Lambda^2+k^2} 
\eeq
with the cut-off momentum, $\Lambda\simeq 1$GeV.
Under the universality ansatz Eq.(25) is reduced into the simple forms,
\beq
\Pi_N^P\rightarrow
\frac{-k^2U^{(0)}_N}{1+g'(U^{(0)}_N+U_\Delta^{(0)})}, \quad
\Pi_\Delta^P\rightarrow
\frac{-k^2U^{(0)}_\Delta}{1+g'(U^{(0)}_N+U_\Delta^{(0)})}
\eeq
with $g'_{NN}=g'_{N\Delta}=g'_{\Delta\Delta}\equiv g'$,
respectively. 

\begin{figure}[h]
\begin{minipage}{0.49\textwidth}
\epsfile{file=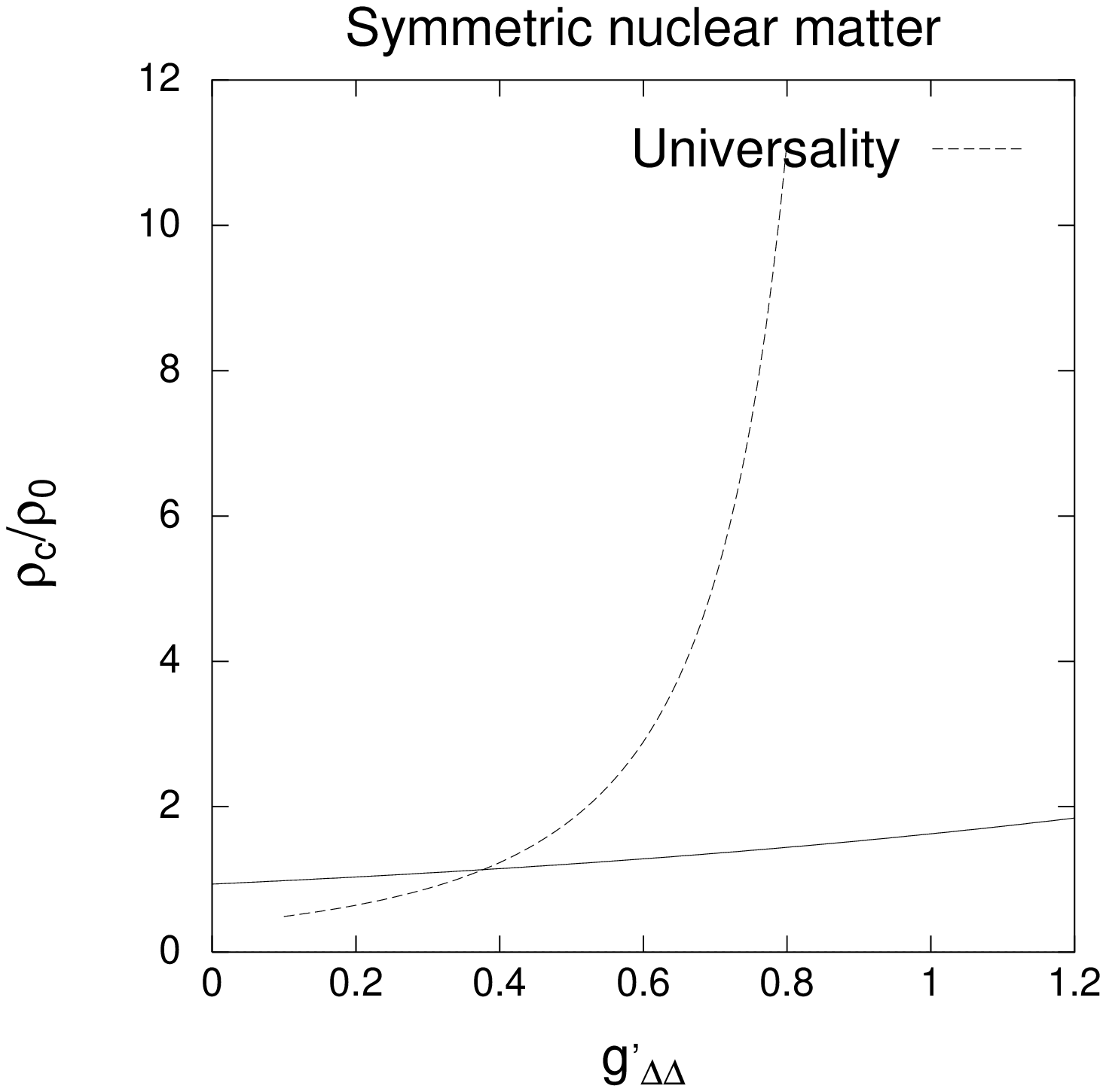,width=\textwidth}
\caption{Critical density for $\pi^0$ condensation in symmetric
nuclear matter as a function of $g'_{\Delta\Delta}$. Dased line shows
the previous one by the use of univesality ansatz. We take $m^*=0.8m$
for the effective mass of nucleons.}
\end{minipage}
\hfill
\begin{minipage}{0.49\textwidth}
\epsfile{file=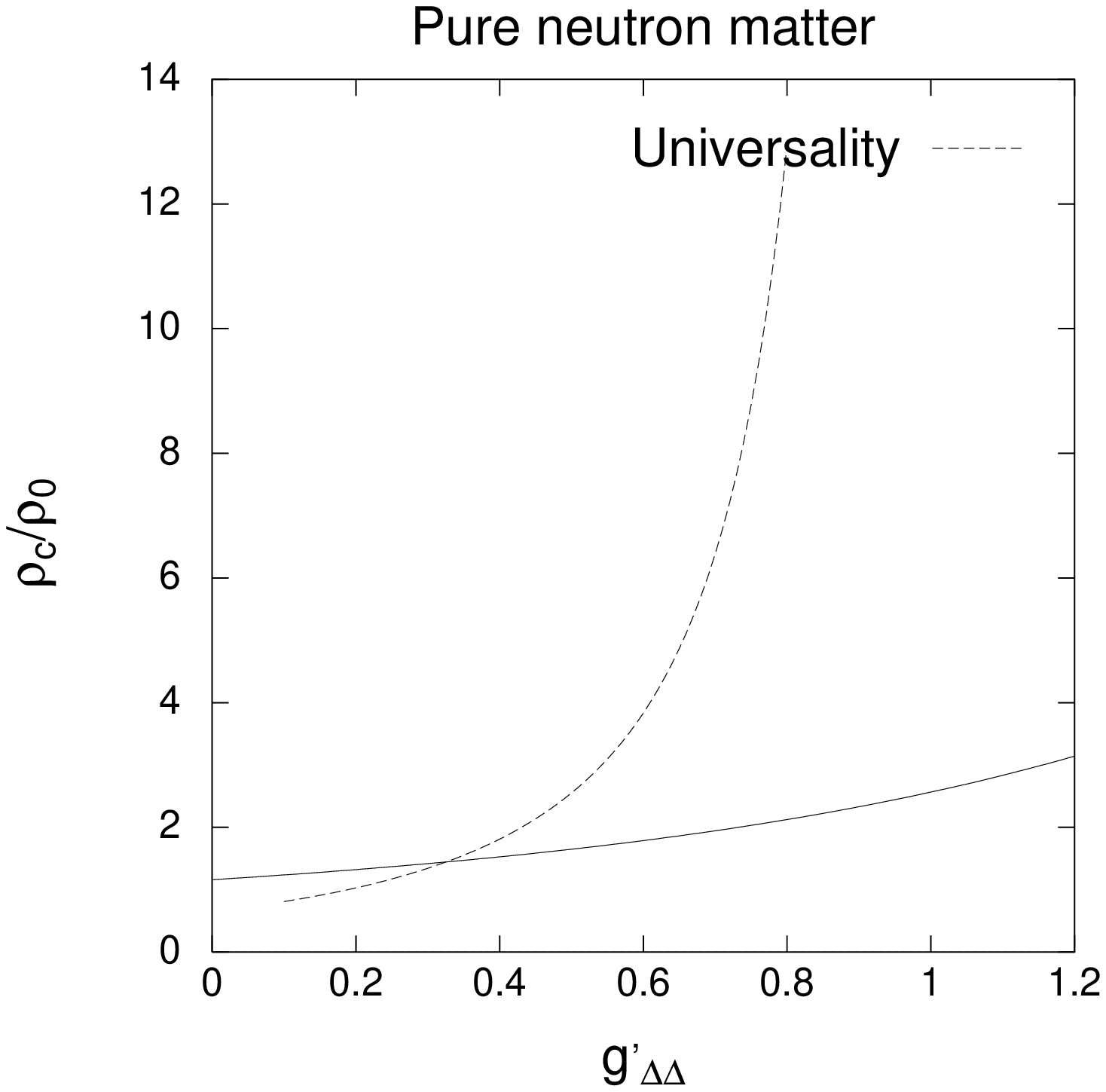,width=\textwidth}
\caption{Critical density for $\pi^0$ condensation in pure neutron
matter as a function of $g'_{\Delta\Delta}$. The meaning of the dashed 
line is the same as in Fig. 1.}
\end{minipage}
\end{figure}

In Figs.1 and 2 we present the critical densities as the 
functions of $g'_{\Delta\Delta}$. We take here $m^*=0.8m$ for the
effective mass of nucleons.
In the previous results with the
universality, the critical density steeply increases at
$g'\sim 0.8$. It is easily understood by Eqs.~(23) and (29); since the
Lindhard functions diverge for $\rho\rightarrow\infty$, the critical
density goes to infinity when the relation,
\beq
  k_c^2=\frac{g'}{1-g'}m_\pi^2
\eeq
holds. Numerically, $k_c=O(m_\pi)$, so that Eq.~(30) may be satisfied
for some value in the range, $0.5<g'<1$.
On the contrary, 
we can see that the $g'_{\Delta\Delta}$ dependence of the critical
density is mild and the singular behavior 
disappears when the universality ansatz is
relaxed. Then, the critical densities result in very {\it low} values:
around $1.6\rho_0$ and $2.5\rho_0$
, for $g'_{\Delta\Delta}=1$, 
in the cases of symmetric nuclear matter and pure neutron matter, respectively.
It might be interesting to compare these values with those indicated
in a recent variational calculation of nuclear matter 
with a modern potential \cite{pan}; $\rho_c\sim 2\rho_0$ and $\sim
1.3\rho_0$ for symmetric nuclear matter and pure neutron matter, respectively.

\begin{figure}[h]
\begin{center}
\epsfile{file=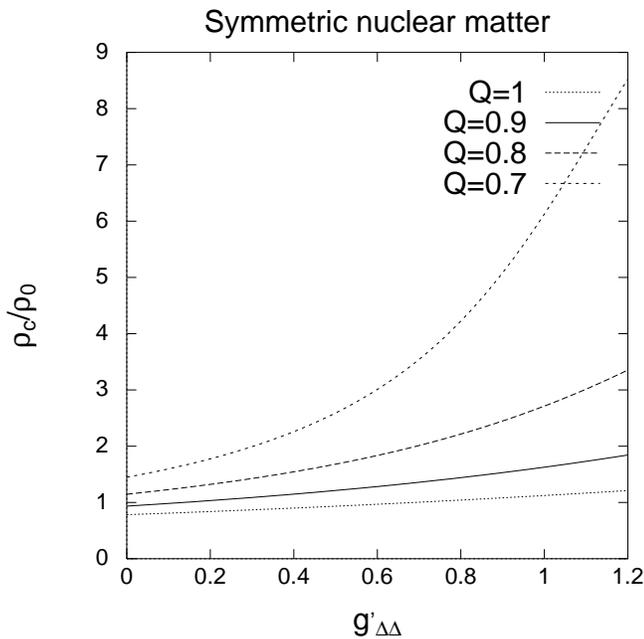,width=0.7\textwidth}
\end{center}
\vspace*{5mm}
\caption{$Q$-dependence of the critical density in symmetric nuclear matter.}
\end{figure}

In. Fig.~3 we show the dependence of the critical density on the value
of $Q$ for the range, 
$0.7<Q<1$. From Eq.~(16) $g'_{N\Delta}$ can be represented,
\beq
g'_{N\Delta}=\beta_2^{-1}(1-\sqrt{Q})(1+\beta_1g'_{\Delta\Delta}),
\eeq
as a function of $g'_{\Delta\Delta}$ for a given $Q$. The $g'_{\Delta\Delta}$
dependence of the critical density becomes more pronounced as the value of
$Q$ becomes smaller; still it is smaller than $2.7\rho_0$ for
$Q>0.8$ as far as $g'_{\Delta\Delta}<1$.

\subsection{Charged pion $(\pi^c)$ condensation in neutron matter}

In pure neutron matter another type of pion condesnations becomes
possible, which is the softening of both the spin-isospin sound
($\pi^+_s$) and
the $\pi^-$ branches, $(\pi^c)$ condensation. It is well-known
that $(\pi^c)$ condensation should be responsible to the non-standard
cooling scenario in the thermal evolution of neutron stars\cite{r}.
In this case, the condensate has the finite energy $\mu_\pi$ due to the 
charge conservation. Then the self-energy by the pion-baryon
interactions consists of the $s$-wave part (Tomozawa-Weinberg term) 
besides the $p$-wave one. Since $\mu_\pi=O(m_\pi)$, the polarization
functions in Eq.~(27) are well-approximated as 
\beq
U^{(0)}_N\sim
\left(\frac{f_\pi\Gamma_\Lambda}{m_\pi}\right)^2\frac{2}{\omega}\rho
, \quad 
U^{(0)}_{\Delta}=\frac{2}{3}\left(\frac{f_\Delta
\Gamma_\Lambda}{m_\pi}\right)^2\left(\frac{1}{\epsilon_\Delta-\omega}
+\frac{1}{3}\frac{1}{\epsilon_\Delta+\omega}\right)\rho.
\eeq
On the other hand, the (repulsive) $s$-wave part is simply given by
\beq
\Pi_N^S({\bf k},\omega;\rho)=\frac{\omega}{2f_\pi^2}\rho.
\eeq
Hence the total self-energy term can be written as
\beq
\Pi=\Pi_N^S+\Pi_N^P+\Pi_\Delta^P.
\eeq

\begin{figure}[h]
\begin{center}
\epsfile{file=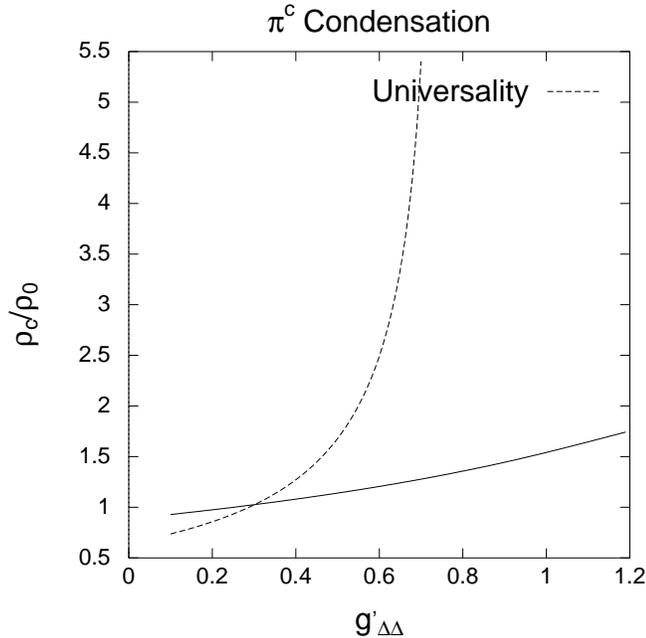,width=0.7\textwidth}
\end{center}
\vspace*{5mm}
\caption{Critical density for $(\pi^c)$ condensation in neutron matter 
as a function of $g'_{\Delta\Delta}$. The meaning of the dashed line
is the same as in Fig. 1.}
\end{figure}

The critical density is given in Fig.~4. It is to be noted that the
dependence of the critical density on the effective mass is very 
weak in this case because of Eq.(32).
The peculiar behaviour, seen in the case of universality, disappears in
our case as for neutral pion condensation. The critical density is
estimated as $1.7\rho_0$ for $g'_{\Delta\Delta}=1$.

\section{Conclusions}

According to the recent data on the quenching factor of the GT strength, we conclude that 
the universality ansatz of the Landau-Migdal parameters does not hold.
The value of $g'_{{\rm N}\Delta}$ is about 0.2, while $g'_{{\rm NN}}$ about 
0.59 in assuming $g'_{\Delta\Delta} < 1.0$. 
Because of the small value of $g'_{{\rm N}\Delta}$, the critical density of the pion condensations weakly depend on the values of $g'_{{\rm NN}}$ and $g'_{\Delta\Delta}$. When the nucleon effective mass is $m^*=0.8m$, the critical density in symmetric nuclear matter is about 1.6$\rho_0$ for $g'_{\Delta\Delta}=1.0$. It is much lower than that predicted before using the universality ansatz.
In neutron matter, the critical density for $\pi^0$ condensation is about 2.5$\rho_0$, while for $\pi^c$ condensation 1.7$\rho_0$, assuming $g'_{\Delta\Delta}$=1.0. The $g'_{\Delta\Delta}$-dependence of the critical densities is weak.
It is interesting to re-examine the possibility of the pion condensations in heavy ion reactions and neutron stars.

\end{document}